\documentclass[%
 reprint,
superscriptaddress,
 amsmath,amssymb,
pra,
]{revtex4-1}

\usepackage{graphicx}
\usepackage{dcolumn}
\usepackage{bm}
\usepackage{hyperref}
\usepackage{makecell}
\usepackage{relsize}

\def\ket#1{|#1\rangle}

\def\aop{\hat{a}}
\def\adop{\hat{a}^\dagger}
\def\dop{\hat{d}}
\def\ddop{\hat{d}^\dagger}
\def\sigop{\hat{\sigma}^-}
\def\sigdop{\hat{\sigma}^+}
\def\Hop{\hat{H}}

\def\tp{{t^\prime}}
\def\kappap{{\kappa^\prime}}

\def\ctil{\tilde{c}}
\def\nn{\nonumber}

\newcommand{\lsz}{\left[}
\newcommand{\rsz}{\right]}
\newcommand{\lk}{\left(}
\newcommand{\rk}{\right)}
\newcommand{\lka}{\left\{}
\newcommand{\rka}{\right\}}

\usepackage{xcolor}

\def\refeq#1{{\hyperref[#1]{(\ref*{#1})}}}
\def\reffig#1{{\hyperref[#1]{FIG. \ref*{#1}}}}
\def\refsec#1{{\hyperref[#1]{SEC. \ref*{#1}}}}
\def\refno#1{{\hyperref[#1]{\ref*{#1}}}}

\begin{document}

\preprint{APS/123-QED}

\title{Comparison between continuous- and discrete-mode coherent feedback for the Jaynes-Cummings model}

\author{Nikolett N\'emet}
\email{nnem614@aucklanduni.ac.nz} 
\affiliation{The Dodd-Walls Centre for Photonic and Quantum Technologies, New
Zealand}
\affiliation{Department of Physics, 
             University of Auckland, Auckland, New Zealand}
\author{Alexander Carmele}%
\email{alexander.carmele@win.tu-berlin.de}
 \affiliation{Nichtlineare Optik und Quantenelektronik, Institut f\"ur 
Theoretische Physik, Technische Universit\"at Berlin, Germany}

\author{Scott Parkins}
\email{s.parkins@auckland.ac.nz}
\affiliation{The Dodd-Walls Centre for Photonic and Quantum Technologies, New
Zealand}
\affiliation{Department of Physics, 
             University of Auckland, Auckland, New Zealand}
             
\author{Andreas Knorr}
\email{andreas.knorr@tu-berlin.de}
\affiliation{Nichtlineare Optik und Quantenelektronik, Institut f\"ur 
Theoretische Physik, Technische Universit\"at Berlin, Germany}

\date{\today}

\begin{abstract}
Using the example of the Jaynes-Cummings model, we present a comparison between time-delayed coherent feedback mediated by reservoirs with continuous and discrete mode structures and work out their qualitative differences. In contrast to the discrete-mode case, the continuous-mode case results in the well-known single-delay dynamics which can, e.g., stabilize Rabi oscillations. The discrete-mode case, however, shows population trapping, not present in the continuous-mode model. Given these differences, we discuss the cavity output spectra and show how these characteristic properties are spectrally identifiable. This work demonstrates the fundamental difference between the continuous-mode case, which represents a truly dissipative mechanism, and the discrete-mode case that is in principle based on a coherent excitation exchange process. \ \\

\begin{description}
\item[PACS numbers]
May be entered using the \verb+\pacs{#1}+ command.
\end{description}
\end{abstract}

\pacs{Valid PACS appear here}
\maketitle

\section{Introduction}
Quantum control methods have recently infiltrated a wide range of research fields, including, but not limited to, precision measurement, state preparation, dynamical stabilization, quantum communication and computation, etc. The efficiency of the proposed schemes are usually system-dependent \cite{Wiseman2009,Dong2010,Nielsen2010}. Thus the example of changing a crucial parameter to shift or to extend a phenomenon for different frequencies requires the alteration of the whole setup. However, typical restrictions associated with noise and signal loss are also present due to the surrounding environment and measurement apparatus. To overcome these obstacles, in setups where coherence preservation is not essential, a generalization of classical electronic feedback schemes can be used for controlling the system's behaviour. However, in the case of these measurement-based feedback methods, back-action noise and extra processing delays can significantly degrade the performance of such control schemes \cite{Lloyd2000,Serafini2012,Chen2013,James2014,Zhang2017}.

Rapidly evolving experimental conditions have recently enabled another, coherent feedback scheme, where the measurement step is omitted and the quantum coherent output of a given system is directly fed back into an input channel. If the propagation time spent between the emission and reabsorption is taken into account, the time non-local reservoir-system interaction results in a non-Markovian dynamics for the system \cite{Whalen2015,Whalen2016}. A fundamental example of this is a two-level system in front of a mirror or, in other words, the half-cavity setup.  In this case, the presence of the mirror imposes a coherent time-delayed feedback for the atom and thus alters the quantum statistics and spontaneous emission spectrum of the system  \cite{Dorner2002,Beige2002,Glaetzle2010,Tufarelli2014,Guimond2017,Faulstich2018,Fang2018,Calaj2018,Fang2019}. Signatures of non-Markovian behaviour for this setup were also demonstrated experimentally \cite{Eschner2001,Wilson2003,Dubin2007}.

Coherent feedback for more complex systems was first introduced under the name of all-optical feedback by Wiseman and Milburn \cite{Wiseman1994}. Due to the coherence-preserving nature of this method, it is much more successful in controlling quantum systems than its measurement-based counterpart \cite{Jacobs2014,Horowitz2014,Yamamoto2014,Roy2017,Kashiwamura2018,Jacobs2015}. Since then, experiments using coherent feedback verified its potential to enhance the efficiency of intrinsic quantum processes \cite{Nelson2000,Iida2012,Zhou2015,Wang2015,Guo2017b}, tune the coupling between different system components \cite{Kerckhoff2013}, alter the stability landscape of the whole quantum system \cite{Mabuchi2008,Kerckhoff2012}, and implement quantum computation tasks \cite{Hirose2016}. 

Experimental successes were matched by a substantial number of theoretical proposals, where the feedback environment was always assumed to have a continuous-mode structure. As pointed out in \cite{Wiseman1994}, such a scheme has to ensure, e.g., by unidirectional propagation, that standing wave modes cannot build up in the feedback loop. Initially, the time delay corresponding to the feedback loop was considered only as a practically unavoidable small influence which, due to the high propagation speed, can be neglected, leading to Markovian dynamics \cite{Gough2009,Iida2012,Zhang2012a,Liu2013,Joshi2014,Pavlichin2014,Yang2015,Nurdin2015,Dong2016,Pan2016,Li2017,Kashiwamura2017,Guo2017b,Xue2017}. However, a long enough feedback loop enforces non-trivial time evolution, where the time delay becomes an important control parameter \cite{Grimsmo2014,Hein2014,Hein2015,Kopylov2015,Kabuss2015,Joshi2016,Guimond2016,Kraft2016,Nemet2016,Alvarez-Rodriguez2017a,Guo2017a,Lu2017,Guimond2017,Pichler2017,Chang2018,Fang2018,Calaj2018,Droenner2018,Nemet2018}.

The effect of a feedback scheme can be entirely different, if instead of a continuous-mode spectrum, a discrete set of modes is considered for the feedback reservoir, as can be the case for the setups analyzed in \cite{Mabuchi2008,Yan2011,Hamerly2012,Hamerly2013,Zhou2015,Wang2015,Yamamoto2016,Wang2017,Chalabi2018}.
To demonstrate these fundamental differences, this paper presents a comparison between the properties of the continuous- and discrete-mode schemes through the example of the Jaynes-Cummings model in the single excitation limit. Each setup has its advantages and disadvantages to be taken into account when constructing more complicated control schemes, such as plant-controller \cite{Mabuchi2011,Zhang2011,Xue2012,Bian2012,Zhang2012b,Crisafulli2013,Vy2013,Albertini2013,Emary2014,Shi2014,Shi2015b,Liu2015,Yokotera2016,Sarovar2016,Thethi2017,Vuglar2017,Balouchi2017,Cui2017,Yan2017,Nguyen2017,Yoshimura2018,Zhang2018} or quantum network systems \cite{Gough2008,Zhang2013,Bennett2014,Shi2015a,Gough2016,Tabak2016,Nurdin2016,Neto2017,Combes2017,Nurdin2017,Lubasch2018,Tabak2018}. This work also further refines the scope of the existing numerical methods that were invented for the non-trivial task of simulating the non-linear time evolution of a quantum system with time non-local interactions \cite{Grimsmo2015,Pichler2016,Whalen2017,Chalabi2018}.

In this Paper, we first introduce in Section II the two models that we consider for implementing time-delayed feedback. The main difference lies in the mode structure of the feedback reservoir. Multiple works have discussed the continuous-mode (CM) case, where the emitted excitation can be lost in the continuum of modes \cite{Carmele2013,Kabuss2015,Chang2018}, however Rabi oscillations can still be recovered. Considering discrete modes for the reservoir, which are determined by the time delay, gives an opportunity to model large, multimode systems in the context of coherent feedback. This is especially interesting in the context of quantum networks, where two nodes such as, e.g., cavity quantum electrodynamics systems, are coupled by a long fibre that has multiple, closely spaced modes \cite{Kato2019}.

Subsequently, in Section III, we derive the time-evolution of the excitation probabilities and compare them in the limiting cases of very short and very long delay. In Section IV we discuss the stability landscape of the two setups, where we demonstrate that although in the continuous-mode case Rabi oscillations can be stabilized, no such claim can be made in the discrete-mode case. On the other hand, Section V investigates the example of excitation trapping in the two-level system, that is only present in the discrete-mode case and not in the continuous-mode case and can be explained with single-mode theory. Finally, in order to further emphasize the key differences in a detectable fashion in Section VI, we present the spontaneous emission spectra for both cases via an additional output channel.

\section{System-environment interaction}
\begin{figure}[h!]
\includegraphics[width=0.42\textwidth,trim={12cm 0 0 0},clip]{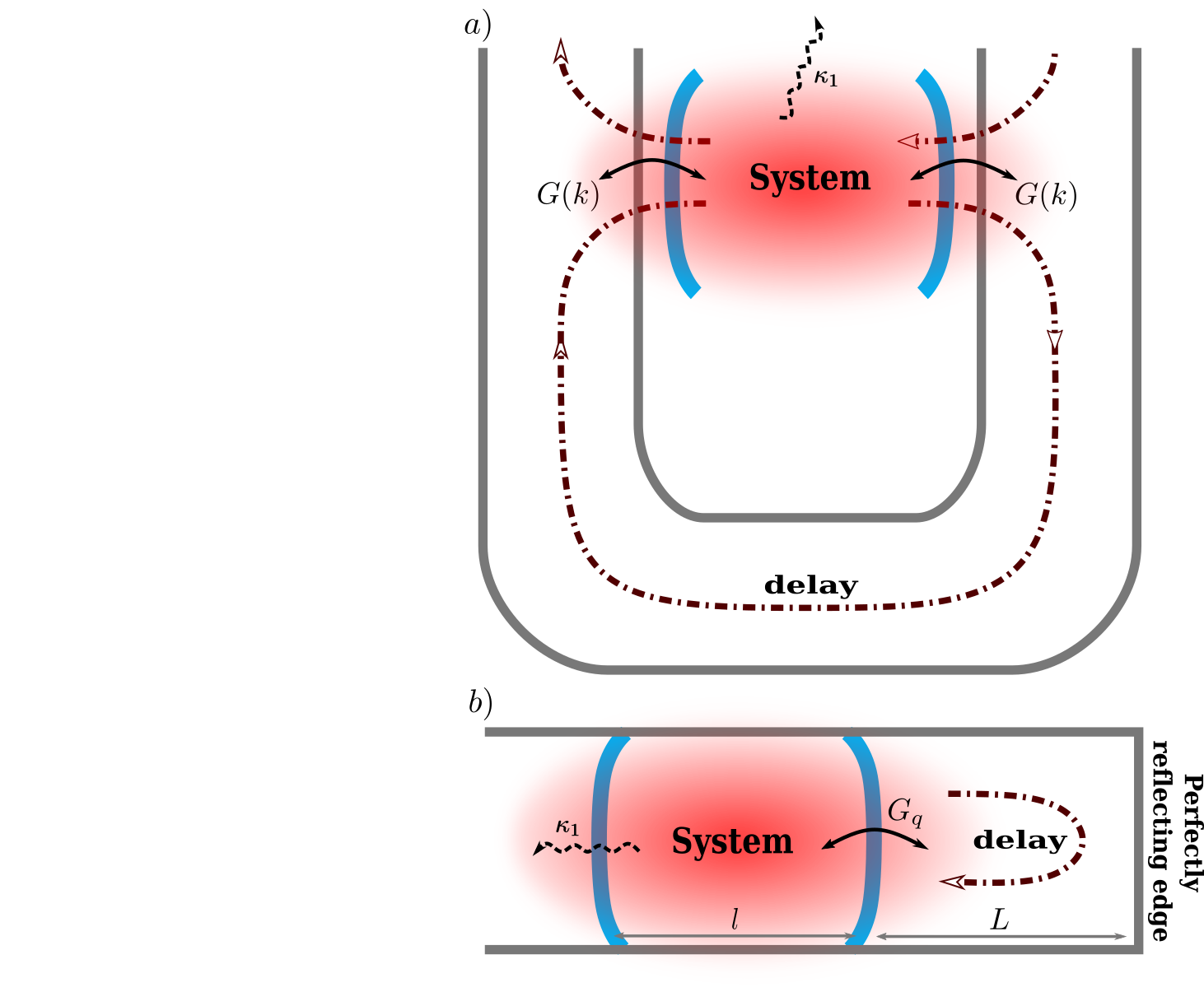}%
\vspace{-.4cm}
\textit{\textbf{\caption{\linespread{1}\small{Coherent time-delayed feedback types for a System bounded by cavity-mirrors. a) In the continuous-mode case the feedback occurs due to the coupling of the system to the reservoir at two different spatial points. Here the reservoir is not confined in any sense, therefore by considering travelling wave modes, the two-point interaction translates into a two-time interaction between the system and the environment. This enforces non-Markovian dynamics with a single delay. b) In the other, discrete-mode scenario the feedback occurs due to the perfectly reflecting mirror or edge of a waveguide. In this case the reservoir is confined into a finite space, which results in a discrete set of modes. This can also be interpreted as a coherent feedback signal with delays as multiples of the returning time to the System.
\label{fig:setup}}}}}
\end{figure}

In this paper we focus on systems that are bounded by cavity mirrors and time non-local system-reservoir interactions. Although both schemes in \reffig{fig:setup} impose coherent feedback on the system, the difference between the mode structure of the reservoirs results in qualitatively different system-reservoir interactions. This, in return, creates distinct time evolution and output characteristics of the system in the two cases.

\subsection{Hamiltonian}
Let us first focus only on the specifics of the feedback interaction keeping the description of the system as general as possible. Our Hamiltonian can be written as
\begin{align}
\Hop_1 &= \Hop_S +\Hop_{S-C} + \hbar\omega_c\adop\aop,
\end{align}
where $\Hop_S$ describes the system dynamics, $\Hop_{S-C}$ characterizes the interaction between system variables and the cavity, and the last term describes the free evolution of the cavity enclosing our system. 

This setup interacts with a discrete- (DM) or continuous-mode (CM) environment via the cavity, which can be described in the interaction picture (with $\Hop_0=\Hop_S+\hbar\omega_c\adop\aop$) by
\begin{align}
\label{eq:Ham_DM}
\Hop^{(DM)}_{B-S} &= -\sqrt{\frac{\pi}{2L}}\sum_{q=-\infty}^\infty\lk\hbar G_{q}(t)\adop\dop_{q} + \hbar G^*_{q}(t)\ddop_{q}\aop\rk,\\
\label{eq:Ham_CM}
\Hop^{(CM)}_{B-S} &= -\int_{-\infty}^\infty\lk\hbar G(k,t)\adop\dop_k + \hbar G^*(k,t)\ddop_k\aop\rk dk,
\end{align}
where $\dop_q$ and $\dop_k$ are discrete and continuous-mode operators, respectively. The factor $\sqrt{\pi/2L}$ comes from the difference between the discrete and continuous-mode description \cite{Blow1990}. For discrete modes the mode spacing in $k$-space is taken to be $\frac{\pi}{L}$, showing a strong connection between the quantization volume and the feedback delay. In the case of continuous modes, on the other hand, the quantization is completely independent of the feedback roundtrip time $2L/c$. $G_q$ and $G(k)$ describe the dispersive coupling between the reservoir modes and the cavity as follows:
\begin{align}
    \label{eq:G_disc}
    G_q(t) &= G_0\sin{(k_qL)}e^{-i(\omega_q-\Delta_0)t},\\
    \label{eq:G_cont}
    G(k,t) &= G_0\sin{(kL)}e^{-i(\omega-\Delta_0)t},
\end{align}
where $\Delta_0$ describes the detuning between the central mode and the cavity resonance and $\omega_q=ck_q$, with wave propagation speed $c$.

The treatment of the discrete-mode setup is analogous to the modes of the universe approach \cite{Lang1973} without taking the limit of infinitesimal spacing, as $L$ characterizes the size of the quantization volume. In the continuous-mode case one can quantize the reservoir modes around the cavity and then take the size of the box to infinity \cite{Whalen2015}. 

In both cases the feedback length is much larger than that of the cavity $l\ll L$. This allows for the limits in the sum and integration to go to $\pm\infty$. The wave number of the discrete set of modes has the expression \cite{Lang1973}:
\begin{align}
\label{eq:kL}
k_qL=\frac{(2q+1)\pi}{2}+\theta_q,
\end{align}
where $\theta_q\ll1$.

A time delay associated with a single roundtrip can be defined as:
\begin{align}
\label{eq:tau}
\tau = \frac{2L}{c},
\end{align}
which is the time between consecutive interactions of the field emitted into the reservoir with the system.
\subsection{Example: Jaynes-Cummings model}
In order to demonstrate the differences between the two feedback schemes, we assume only a single, initially excited atom in the cavity. Thus the system is described by the Jaynes-Cummings interaction Hamiltonian in the rotating-wave approximation, and the cavity field frequency is tuned to the atomic resonance.
The total Hamiltonian guiding the evolution of the two-level system, the cavity and the reservoir is described by
\begin{align}
\label{eq:Ham}
    \Hop = \hbar\gamma\lk\sigdop\aop+\adop\sigop\rk -\Hop_{B-S}^{(CM/DM)}
\end{align}
for the continuous- and discrete-mode case, respectively.

In the following sections we provide a systematic overview of the differences between these two cases in various parameter regimes labelled by two dimensionless parameters, $\kappa\tau$ and $\gamma/\kappa$.
The delay parameter, $\kappa\tau$, describes the width of the emitted wave packet in the time domain compared to the length of the roundtrip time in the feedback loop. In the short delay case $(\kappa\tau\ll1)$ the time it takes for the cavity to completely decay to its ground state is longer than the feedback time, therefore the output field interferes with itself. On the other hand, for a substantially long time delay in the feedback loop $(\kappa\tau\gg1)$, every roundtrip represents an independent input for the system. 

The coupling parameter, $\gamma/\kappa$ determines the relative strength of the coupling between the system and the cavity compared to the influence of the feedback loop. In the strong coupling regime ($\gamma/\kappa\gg1$) the dynamics is dominated by the coherent exchange of excitation between the resonator and the TLS. In the weak coupling regime ($\gamma/\kappa\approx1$) there is a competition between the influence of the environment and that of the atom on the cavity field. In the bad cavity regime ($\gamma/\kappa\ll1$) the time evolution is mainly steered by the feedback dynamics.

\section{Single excitation limit}
We assume an initially excited atom and no external driving for the cavity. The number of excitations is kept constant, which results in a linear dynamics.

\subsection{Continuous-mode case}
We start with briefly reviewing the dynamics of the continuous-mode case as presented in \cite{Kabuss2015} for comparison. The single-excitation wave function is presented as the linear combination of atomic, cavity and reservoir excitations with different weights.
\begin{align}
\ket{\psi(t)^{(CM)}} &= c_e^{(CM)}(t)\ket{e,0,\{0\}} + c_g^{(CM)}(t)\ket{g,1,\{0\}}  \nn\\
&\hspace{.4cm}+\int c_{g,k}^{(CM)}(t)\ket{g,0,\{k\}}dk.
\end{align}
Using this wave function in the Schr\"odinger equation with Hamiltonians (\ref{eq:Ham_CM}) and (\ref{eq:Ham}), the coefficients follow the time-evolutions
\begin{align}
\label{eq:c_e_eq_SD}
\frac{dc_e^{(CM)}}{dt} &= i\gamma c_g^{(CM)}(t),\\
\label{eq:c_g_eq_SD}
\frac{dc^{(CM)}_g}{dt} &= i\gamma c^{(CM)}_e(t) + i\int G(k,t)c^{(CM)}_{g,k}(t)dk,\\
\label{eq:c_gqp_eq_SD}
\frac{dc^{(CM)}_{g,k}}{dt} &= iG^*(k,t)c^{(CM)}_g(t).
\end{align}
Substituting \refeq{eq:G_cont} into these equations, the system only interacts with one of its past versions via the environment in a Pyragas feedback form \cite{Pyragas1992}:
\begin{align}
\label{eq:CM_EqM}
\frac{dc_g^{(CM)}}{dt} &= i\gamma c_e^{(CM)}(t) - 2\kappa\lsz c_g^{(CM)}(t)\right.\\ &\left.\hspace{3cm}-e^{i\phi}c_{g}^{(CM)}(t-\tau)\Theta(t-\tau)\rsz,\nn
\end{align}

There is a characteristic phase associated with the propagation of the central mode $\phi=\Delta_0\tau$, which prescribes the effect of the environment. The interference between the present and past field of the cavity with phase $\phi=2n\pi$ can enhance the decay of the cavity field, whereas $\phi=(2n+1)\pi$ on the other hand relates to a suppressed decay $(n\in\mathbb{N})$.

\subsection{Discrete-mode case}
The wave function can be expressed similarly to the continuous-mode case in the single-excitation limit:
\begin{align}
\label{eq:ce_EoM}
\ket{\psi(t)^{(DM)}} &= c_e^{(DM)}(t)\ket{e,0,\{0\}} + c_g^{(DM)}(t)\ket{g,1,\{0\}} + \nn\\
&\hspace{.4cm}+\sum_q c_{g,q}^{(DM)}(t)\ket{g,0,\{k_q\}}.
\end{align}
Using the Hamiltonians (\ref{eq:Ham_DM}) and (\ref{eq:Ham}) in the Schr\"odinger equation, we obtain the following time-local equations of motion:
\begin{align}
\label{eq:c_e_eq_MD}
\frac{dc_e^{(DM)}}{dt} &= i\gamma c_g^{(DM)}(t),\\
\label{eq:c_g_eq_MD}
\frac{dc_g^{(DM)}}{dt} &= i\gamma c_e^{(DM)}(t) \nn\\ &\hspace{.4cm}+i\sqrt{\frac{\pi}{2L}}\sum_{q=-\infty}^{\infty}G_q(t)c_{g,q}^{(DM)}(t),\\
\label{eq:c_gqp_eq_MD}
\frac{dc_{g,q}^{(DM)}}{dt} &= i\sqrt{\frac{\pi}{2L}}G^*_q(t)c_g^{(DM)}(t).
\end{align}

Substituting the formal integral of \refeq{eq:c_gqp_eq_MD} into \refeq{eq:c_g_eq_MD} and using \refeq{eq:G_disc} gives:
\begin{align}
\frac{dc_g^{(DM)}}{dt} &= i\gamma c^{(DM)}_e(t) - \frac{|G_0|^2\pi}{2L}\sin^2{(k_qL)}\nn\\
&\hspace{.4cm}\cdot\sum_{q=-\infty}^{\infty}
\int_0^te^{-i(\omega_q+\Delta_0) (t-\tp)} c^{(DM)}_{g}(\tp)d\tp\nn.
\end{align}
The sine factor in the expression above gives 1 as the modes are considered around resonance (see in Appendix \ref{app:sinkL}). After exchanging the summation with the integral, we have the following expression for the sum in $q$:
\begin{align}
&\sum_{q=-\infty}^{\infty}e^{-i(\omega_q+\Delta_0) (t-\tp)} \nn\\
&\hspace{.4cm}= e^{-i\lk\Delta_0+\frac{\pi}{\tau}\rk (t-\tp)}\sum_{q=-\infty}^{\infty}e^{-iq 2\pi \frac{t-\tp}{\tau}} \nn\\
&\hspace{.4cm}=\tau e^{-i\lk\Delta_0+\frac{\pi}{\tau}\rk (t-\tp)}\sum_{q = -\infty}^\infty\delta\lk t-\tp-q\tau\rk,
\end{align}
where we have used the Fourier series identity for the Dirac comb \cite{Strichartz2003} with $\omega_q = c\frac{(2q+1)\pi}{2L}=\frac{(2q+1)\pi}{\tau}$. Thus the equation of motion for $c_g^{(DM)}$ can be expressed as:
\begin{align}
\frac{dc_g^{(DM)}}{dt} &= i\gamma c_e^{(DM)}(t) -\frac{|G_0|^2\pi\tau}{2L}\mathlarger{\mathlarger{\int}}_0^t \Bigg[ e^{-i\lk\Delta_0+\frac{\pi}{\tau}\rk(t-\tp)}\nn\\
&\hspace{.4cm}\cdot
\sum_{q=-\infty}^{\infty}\delta\lk t-\tp-q\tau\rk  c_{g}^{(DM)}(\tp)\Bigg] d\tp.\nn
\end{align}
The range of integration excludes the non-causal values of $q<0$. The vanishing argument of the Dirac $\delta$-function at the upper limit of the integral results in a factor of $1/2$ for $q=0$, thus:
\begin{align}
\label{eq:cg_EoM}
\frac{dc_g^{(DM)}}{dt} &= i\gamma c_e^{(DM)}(t)  \underbrace{\frac{|G_0|^2\pi}{c_0}}_{4\kappa}\cdot\Bigg[\frac{1}{2}c_g^{(DM)}(t)\\
&\hspace{0.4cm}-\sum_{q=0}^{\infty}(-1)^qe^{-i\Delta_0q\tau} c_{g}^{(DM)}(t-q\tau)\Theta(t-q\tau)\Bigg],\nn
\end{align}
which means that each roundtrip contributes to the time-evolution with a corresponding multiple of the delay time $\tau$.
Subsequent roundtrips also mean a phase accumulation of multiples of the characteristic phase identified for the continuous-mode case, i.e., $q\Delta_0\tau$.
\begin{figure}[h!]
\centering
\includegraphics[width=0.5\textwidth,trim={0 0 0 0},clip]{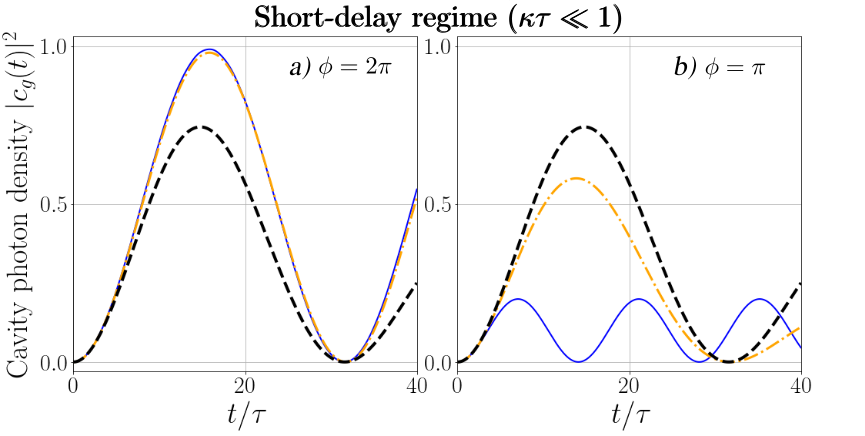}
\vglue -.5 cm
\caption{Time-evolution of the cavity field for a short time delay of $10\kappa\tau=\gamma\tau=0.1$ and a feedback phase of a) $\phi=2\pi$ and b) $\phi=\pi$. The solution without feedback is shown as a black dashed line, whereas the orange dash-dotted line shows the continuous-mode case calculated from \cite{Kabuss2015}, and the blue solid line represents the time-evolution guided by multiple delays.\label{fig:shortdelay}}
\vglue .2 cm
\end{figure}

\subsection{Short-delay limit \texorpdfstring{$(\kappa\tau\ll 1)$}{Lg}}
If the length of the feedback loop is short enough, a Markovian approximation can be made, which means that the propagation time delay is neglected \cite{Wiseman1994}. For the case with a discrete set of modes this translates into the teeth of the Dirac comb merging into a central one.

 For the continuous-mode case, as the delay vanishes, a phase of $\phi=2\pi$ means a reduced effective damping rate according to \refeq{eq:CM_EqM}. The amount of recovered cavity photon probability agrees well between the continuous- and discrete-mode case as shown in \reffig{fig:shortdelay} \textit{(a)}. However, when the feedback phase is set to $\phi=\pi$, the discrete-mode feedback shows different characteristics from the continuous-mode case due to the influence of multiple roundtrips (\reffig{fig:shortdelay} \textit{(b)}).

\begin{figure}[t!]
\centering
\includegraphics[width=0.48\textwidth,trim={0 0 0 0},clip]{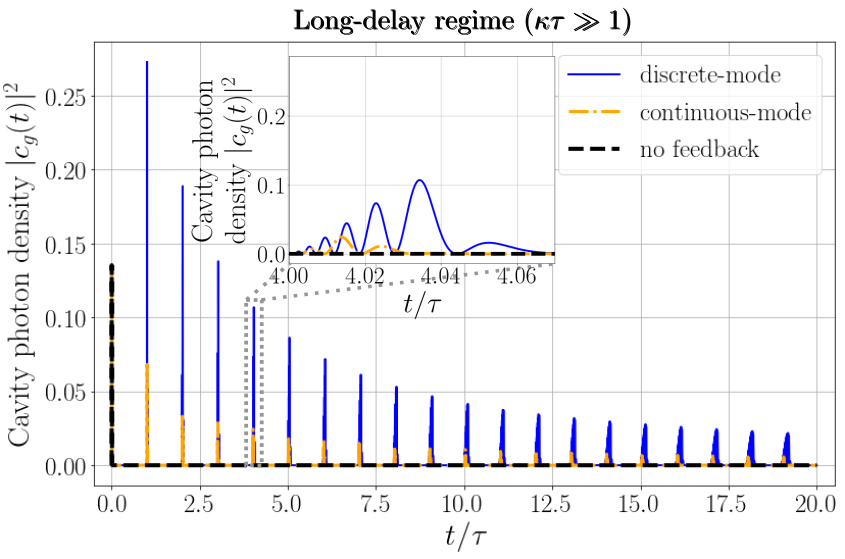}
\vglue -.5 cm
\caption{Time-evolution of the cavity field for a long time delay of $\kappa\tau=\gamma\tau=100\pi,\phi=2\pi$. The continuous-mode case is calculated from \cite{Kabuss2015}.\label{fig:long_delay}}
\vglue .2 cm
\end{figure}

\subsection{Long-delay limit \texorpdfstring{$(\kappa\tau\gg1)$}{Lg}}
As the length of the feedback loop grows, the first roundtrip turns into another Markovian limit, which is exactly the evolution without any feedback considered. For the discrete-mode case this can be interpreted as having so much space between the teeth of the Dirac comb, that only one of them influences the system's evolution. Note how FIGS \ref{fig:long_delay} and \ref{fig:3tau} show this effect as all 3 curves agree before $t/\tau=1$.

In this subsection we compare the two cases in this long-delay limit with a feedback length of $\kappa\tau=100\pi\gg1$. In both cases the field completely leaves the cavity before the front of the wave-packet returns, resulting in a minimal interference with itself.

Focusing only on the large scale evolution of the cavity field in \reffig{fig:long_delay}, both approaches seem to show the same pulsed dynamics. However, zooming in for the individual peaks at multiples of $\tau$ (\reffig{fig:long_delay} inset), the characteristic micro-evolution of the pulses differs significantly between the two models. In the discrete-mode case, each roundtrip adds an extra oscillation, whereas for the continuous-mode case the initial pulse shape is preserved.

After considering the short- and long-delay regimes, in the following we focus on an intermediate regime of $(\kappa\tau\approx1-10)$. As the timescales of the feedback and the cavity decay become comparable, the specific system-environment interaction has a more substantial role in the feedback dynamics.

\section{Special points in the stability landscape}
Previously, stabilized Rabi oscillations were reported in the case of continuous-mode coherent feedback \cite{Kabuss2015}. Such a phenomenon can be identified by a purely imaginary pole in the stability landscape of the dynamical system \cite{scholl2008}. In order to find such a special parameter set in the discrete-mode case, we take the Laplace Transform of the equations of motion derived in the previous section (\ref{eq:ce_EoM},\ref{eq:cg_EoM}):
\begin{align}
\label{eq:lap_ce}
s\ctil_e^{(DM)}(s) &= 1+i\gamma\ctil_g^{(DM)}(s),\\
\label{eq:lap_cg}
s\ctil_g^{(DM)}(s) &= i\gamma\ctil_e^{(DM)}(s) -4\kappa\ctil_g^{(DM)}(s)\nn\\
&\hspace{.5cm}\cdot\lsz\sum_q (-1)^qe^{-q(s+i\Delta_0)\tau}-\frac{1}{2}\rsz.
\end{align}

The general solution at $n\tau\le t<(n+1)\tau$ has the following form:
\begin{align}
\label{eq:DM_laplace}
&\ctil_g^{(n)(DM)}(s) \nn\\
&\vspace{.4cm}=\frac{i\gamma}{s^2+\gamma^2+2\kappa s\lka2\sum_{q=0}^{n}\lsz-e^{-(s+i\Delta_0)\tau}\rsz^q-1\rka}.
\end{align}

This expression is quite different from the continuous-mode solution of
\begin{align}
\label{eq:CM_laplace}
\ctil_g^{(\infty)(CM)}(s) = \frac{i\gamma}{s^2+\gamma^2+2\kappa s\lk1-e^{-(s+i\Delta_0)\tau}\rk}.
\end{align}
Setting the denominator to zero in the general solutions \refeq{eq:DM_laplace} or \refeq{eq:CM_laplace} provides the poles of the stability landscape, which determine the dominant dynamics of the system.

\paragraph{$t<\tau$}
As the general solution for the discrete-mode case \refeq{eq:DM_laplace} is valid between $n\tau$ and $(n+1)\tau$, $t<\tau$ translates as $n=q=0$. This means that both \refeq{eq:DM_laplace} and \refeq{eq:CM_laplace} simplify to the following:
\begin{align}
\ctil_g^{(0)}(s) = \frac{i\gamma}{s^2+\gamma^2+2\kappa s}.
\end{align}
Thus, before the return of the first output pulse both configurations show the usual Markovian dynamics without feedback.
The frequency of the damped oscillations is influenced by the coupling strength between the cavity and the external modes.

\begin{figure}[h!]
\centering
\vglue -.2 cm
\includegraphics[width=0.48\textwidth,trim={0 0 0 0},clip]{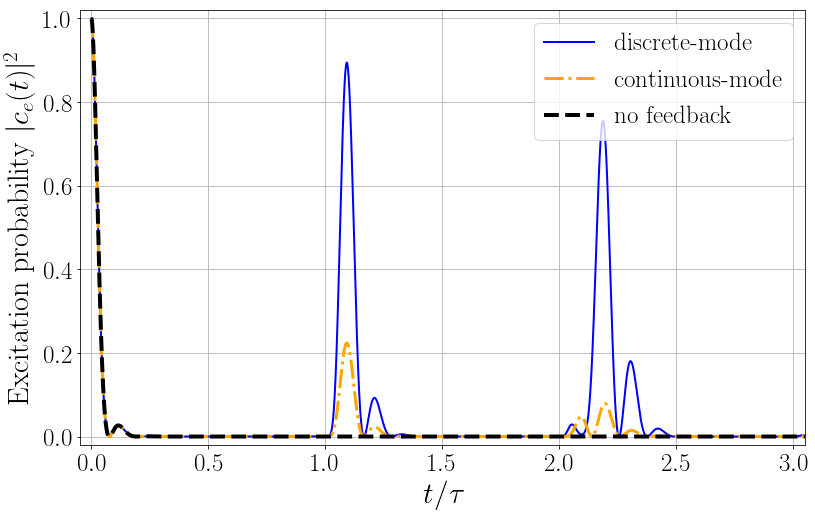}
\vglue -.5 cm
\caption{The dynamics of the discrete- and continuous-mode feedback show clear differences after the second roundtrip. $\kappa\tau = 0.5\gamma\tau = 5\pi, \phi=\pi$. The continuous-mode case is calculated from \cite{Kabuss2015}.\label{fig:3tau}}
\end{figure}

\paragraph{$\tau<t\leq 2\tau$}
In the next time period, after $\tau$, the main dynamical features of the discrete-mode case resembles the continuous-mode case in \refeq{eq:CM_laplace}:
\begin{align}
\label{eq:first_int}
\ctil_g^{(1)(DM)}(s) = \frac{i\gamma}{s^2+\gamma^2+2\kappa s\lk1-2e^{-(s+i\Delta_0)\tau}\rk}.
\end{align}
The main difference in this time period originates from the ratio between the contribution of the recurring field and the original decay, which is half as much in the discrete-mode case as for the continuous-mode feedback. Notice how the main dynamical features of the first two pulses are similar in \reffig{fig:3tau}, but after the second roundtrip they start to deviate from each other significantly.

\paragraph{Long-time limit}
In the continuous-mode case, the equilibrium evolution shows recovered Rabi oscillations (\reffig{fig:trapped} inset) when $(\Delta_0+\gamma)\tau=2n\pi,\  (n\in \mathbb{N})$ \cite{Kabuss2015}:
\begin{align}
    c_g(t) = \frac{i\sin{\gamma t}}{1+\kappa m\pi/\gamma}, \quad m\in\mathbb{N}.
\end{align}

Examining the poles of the discrete-mode case, we assume that there are persistent oscillations and thus, there are such poles as $s_{osc}=\pm i\mu$. 

Substituting this expression into \refeq{eq:DM_laplace} and setting the denominator to 0, the following is obtained:
\begin{align}
\label{eq:stab}
-\mu^2+\gamma^2\mp i2\kappa \mu\lka 1-2\sum_{q=0}^{n}\lsz-e^{\mp i(\mu+\Delta_0)\tau}\rsz^q\rka&=0,
\end{align}
where setting the real and imaginary parts to $0$, respectively, gives
\begin{align}
\label{eq:cos}
\sum_{q=0}^n (-1)^q\cos{\lsz\mp q(\mu+\Delta_0)\tau\rsz} &= \frac{1}{2},\\
-\mu^2+\gamma^2-4\kappa\mu\sum_{q=0}^n (-1)^{q}\sin{\lsz\mp q(\mu+\Delta_0)\tau\rsz}&=0.
\end{align}

For $\mu=\gamma$, reported in \cite{Kabuss2015}, in the long time limit, the sum in \refeq{eq:stab} turns into an infinite geometric series. However the absolute value of the individual terms are 1, which means that the sum does not converge. This contradiction means that there is no such parameter set that would result in recovered Rabi oscillations. Similarly, in general, for $\mu\in\mathbb{R}$ this condition turns into a non-convergent cosine series \refeq{eq:cos}, which means that the effect of each $\tau$ interval depends on the overall phase $(\mu+\Delta_0)\tau$.

\section{Excitation trapping}
Next, we discuss the possibility to trap some of the atomic excitation in the system in spite of the losses present. In order to demonstrate this effect, let us introduce an extra input-output channel for both models that is characterized by the rate $\kappa_1$.

\subsection{No population trapping in the continuous-mode case}
As we are still interested in the single excitation limit, the equation of motion for $c^{(CM)}_g(t)$ changes the following way:
\begin{align}
\frac{dc_g^{(CM)}}{dt} &= i\gamma c_e^{(CM)}(t) - 2\kappa_1c_g^{(CM)}(t)\nn\\
&\hspace{.4cm}-2\kappa\lsz c_g^{(CM)}(t)-e^{i\phi}c_{g}^{(CM)}(t-\tau)\Theta(t-\tau)\rsz.\nn
\end{align}
Let us use the Laplace transform of this equation to determine the steady state value of the coefficient corresponding to the atomic excitation:
\begin{align}
    &\lim_{s\to0}s\tilde{c}_e^{(CM)}(s) =\frac{1+i\gamma\tilde{c}_g^{(CM)}(s)}{s} \nn\\
    &=\lim_{s\to0}\frac{s\lsz s+2\kappa\lk1-\text{exp}{(-i\phi-s\tau)}\rk+2\kappa_1\rsz}{s^2+\gamma^2+2\kappa s\lk1-\text{exp}{(-i\phi-s\tau)}\rk+2\kappa_1s}\nn\\
    &=\frac{0}{\gamma^2}=0.
\end{align}
Thus, no excitation trapping can be observed for a continuous-mode single-delay coherent feedback (\reffig{fig:trapped}).

\subsection{Population trapping in the discrete-mode case}
In this case the extra loss channel can be interpreted as a finite transmission through the mirror on the left-hand side in \reffig{fig:setup} b). The corresponding equation of motion for the cavity field has the following form:

\begin{align}
\frac{dc_g^{(DM)}}{dt} &= i\gamma c_e^{(DM)}(t) - 2\kappa_1c_g^{(DM)}(t)+4\kappa\lsz\frac{1}{2}c_g^{(DM)}(t)-\right.\nn\\
&\left.\hspace{-0.5cm}-\sum_{q=0}^{\infty}(-1)^qe^{-i\Delta_0q\tau} c_{g}^{(DM)}(t-q\tau)\Theta(t-q\tau)\rsz.
\end{align}

\begin{figure}[t!]
\centering
\vglue -.2 cm
\includegraphics[width=0.48\textwidth]{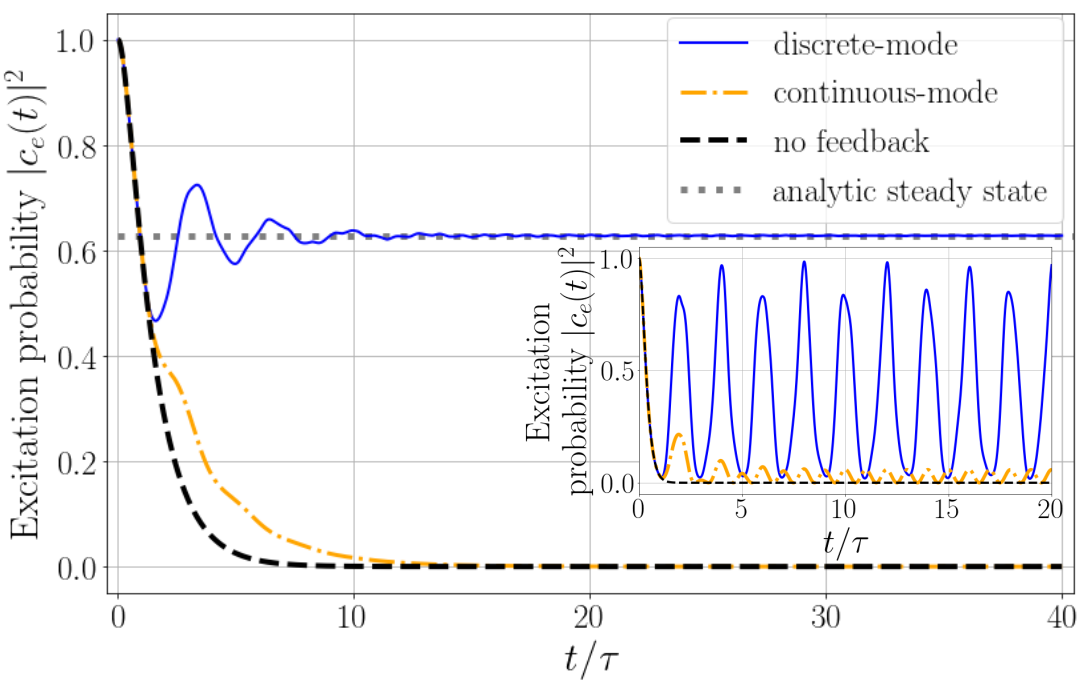}
\vglue -.5 cm
\caption{Time-evolution of the atomic excitation probability for a leaky cavity field with $\kappa\tau=\gamma\tau=\pi/3, 2\kappa_1=\kappa,\phi=\pi$. The continuous-mode case is calculated from \cite{Kabuss2015}. The steady state limit is evaluated using \refeq{eq:SS_trapped}. The inset shows the stabilized Rabi oscillations for a continuous-mode feedback when $\kappa_1=0$ and $\kappa\tau=\pi$.\label{fig:trapped}}
\vglue -.1 cm
\end{figure}

For any open quantum system one would expect that after reaching the steady state all excitations in the system are lost to the surrounding environment, as shown for the continuous-mode case in the previous subsection. However, in the case of discrete-mode coherent feedback, there is a finite probability to find the atom in its excited state (\reffig{fig:trapped}). 
This finite excitation in the steady state can also be shown by following the same procedure as before and look at the Laplace transformed solution:
\begin{align*}
    \tilde{c}_e^{(DM)}(s) &= \frac{1+i\gamma \tilde{c}_g(s)}{s}\nn\\
    &\hspace{-1cm}=\frac{s+4\kappa\lka\sum_{q=0}^\infty\lsz- e^{-(i\phi+s\tau)}\rsz^q-\frac{1}{2}\rka+2\kappa_1}{s^2+\gamma^2+4\kappa s\lka\sum_{q=0}^\infty \lsz-e^{-(i\phi+s\tau)}\rsz^q-\frac{1}{2}\rka+2\kappa_1 s}.
\end{align*}
Assuming the existence of a steady state solution translates as a finite positive real part for $s$. In this case the infinite sum converges, and we find
\begin{align*}
    \tilde{c}_e^{(DM)}(s) &=\frac{s+\frac{4\kappa}{1+\text{exp}{(-i\phi-s\tau)}}-2\kappa+2\kappa_1}{s^2+\gamma^2+\frac{4\kappa s}{1+\text{exp}{(-i\phi-s\tau)}}-2\kappa s+2\kappa_1 s}.
\end{align*}
Using this Laplace transform, the steady state solution can be obtained as:
\begin{align}
    \lim_{s\to0}s\tilde{c}_e^{(DM)}(s)=\frac{\frac{4\kappa s}{1+\text{exp}{(-i\phi-s\tau)}}}{\gamma^2+\frac{4\kappa s}{1+\text{exp}{(-i\phi-s\tau)}}},
\end{align}
where the vanishing terms were omitted. For $\phi=(2n+1)\pi,\ (n\in\mathbb{N})$ the limit of $\frac{4\kappa s}{1-\text{exp}{(-i\phi-s\tau)}}$ is $0/0$. Using L'H\^opital's rule we find:
\begin{align}
    \lim_{s\to0}\frac{4\kappa s}{1-\text{exp}{(-s\tau)}} = \frac{4\kappa}{\tau}.
\end{align}
Thus, the steady-state solution has the following form:
\begin{align}
    \label{eq:SS_trapped}
    \lim_{s\to0}s\tilde{c}_e^{(DM)}(s) &= \frac{1}{1+\eta}, & \eta &= \frac{\gamma^2\tau}{4\kappa}.
\end{align}
This expression is shown as a grey dotted line in \reffig{fig:trapped}. Note that as $\tau\to\infty$, the steady state solution goes to 0, which is the limit of an infinite reservoir, where all system excitations are lost and Markovian time evolution is recovered. Note that the steady state population does not depend on the rate associated with the extra decay channel $\kappa_1$.

In order to have an intuitive picture about where the excitation can "hide" from the losses, we note that the discrete-mode feedback reservoir can also be interpreted as a multimode cavity coupled to the original cavity-QED system. In the following subsection we show that this allows for an approximate description of the process from another perspective where both cavities are treated with a single mode.

\subsection{Single mode theory: Normal modes}

Let us consider two cavities that are directly coupled to each other, with an atom inside cavity 1 on the left. In case of two single-mode cavities with mode frequencies matching the atomic resonance, we can look at the equations of motion in the weak driving limit. Then the following normal modes can be observed in the frame rotating at the atomic resonance frequency:
\begin{align*}
    \ket{B_\pm} &= \frac{1}{\sqrt{2}\xi}\lk \gamma\ket{A} \pm \xi\ket{C_1}+G\ket{C_2}\rk, &    E_{B_\pm} &= \pm\hbar\xi,\\
    \ket{D} &= \frac{1}{\xi}\lk -G\ket{A} +\gamma\ket{C_2}\rk, &
    E_D &= 0, 
\end{align*}
with
\begin{align*}
    \xi&=\sqrt{\gamma^2+G^2},
\end{align*}
where $\ket{C_1}$ and $\ket{C_2}$ represent a coherent excitation for cavity 1 and 2 respectively, and $\ket{A}$ describes the excited state of the atom in the cavity on the left. The emerging normal modes $\ket{B_\pm}$ are bright states and $\ket{D}$ is dark to the leaky cavity. $\gamma$ is the coupling strength between cavity 1 and the atom as before, and $G = \sqrt{\frac{\pi}{c_0}}G_0=2\sqrt{\frac{\kappa}{\tau}}$ describes the interaction strength between the two cavities.

Now let us consider an initially excited atom. This means that we have the following state in terms of the normal modes:
\begin{align}
    \ket{A} = \frac{1}{\xi}\lsz \frac{\gamma}{\sqrt{2}}\lk\ket{B_+}+\ket{B_-}\rk-G\ket{D}\rsz.
\end{align}

As the only decay considered here is that of the left-hand side cavity, an initially excited atom can preserve some of its excitation via the above mentioned dark state $\ket{D}$. All the other state contributions decay away. Thus, we have the following coefficient for the atomic excited state:
\begin{align}
    -\frac{G}{\xi}\left<A\left|\right.D\right>=\frac{G^2}{|\xi|^2},
\end{align}
which is exactly the same as in \refeq{eq:SS_trapped}. Therefore the stronger the cavities are coupled compared to the cavity-atom coupling, the more excitation is preserved in the atom. In the discrete-mode feedback setup this means the bad cavity limit.

\section{Cavity output spectra}
In this section we derive the spontaneous emission spectra through the additional output channel introduced in the previous section $(\kappa_1)$, which enables us to show qualitative differences in an observable manner. We follow the same procedure as in \cite{Zeeb2015} by considering the coefficients in the single excitation limit. In this case the double integral of the two-time correlation function simplifies to
\begin{align}
S(\omega) = \frac{2\kappa_1}{\pi}\left| \ctil_g(-i\omega)\right|^2,
\end{align}
where the Laplace transform of the cavity excitation coefficient is taken at $-i\omega$. 

\paragraph{Discrete-mode feedback}The Laplace transformed solution has the following form for the discrete-mode coherent feedback in the damped case:
\begin{align}
\ctil_g^{(DM)}(s) &= i\gamma\Bigg( s^2+\gamma^2+2\kappa_1s \nn\\
&\hspace{.7cm}- 4\kappa s\lka\sum_{q=0}^\infty \lsz -e^{-(s\tau+i\phi)}\rsz^q-\frac{1}{2}\rka\Bigg)^{-1}, \nn\\
\ctil_g^{(DM)}(-i\omega)&=i\gamma\lka -\omega^2+\gamma^2-i2\kappa_1\omega \right.\nn\\
&\hspace{1cm}\left.- 2\kappa\omega\tan{\lsz\lk\omega\tau-\phi\rk/2\rsz}\rka^{-1},
\end{align}
where assuming a steady state the sum in this expression is convergent. 
 Thus the spontaneous emission spectra has the following form:
\begin{align}
    S^{(DM)}(\omega) &= \frac{2\gamma^2\kappa_1}{\pi}\Bigg\{\bigg[\omega^2-\gamma^2\\
    &\hspace{1.3cm}+2\kappa\omega\tan{\lk\frac{\omega\tau-\phi}{2}\rk}\bigg]^2+
    4\kappa^2_1\omega^2\Bigg\}^{-1}\nn.
\end{align}

Similar expressions can be derived for the case without feedback as well as for the continuous-mode feedback example.

\begin{figure}[t!]
\centering
\includegraphics[width=0.48\textwidth,trim={0 0 0 0},clip]{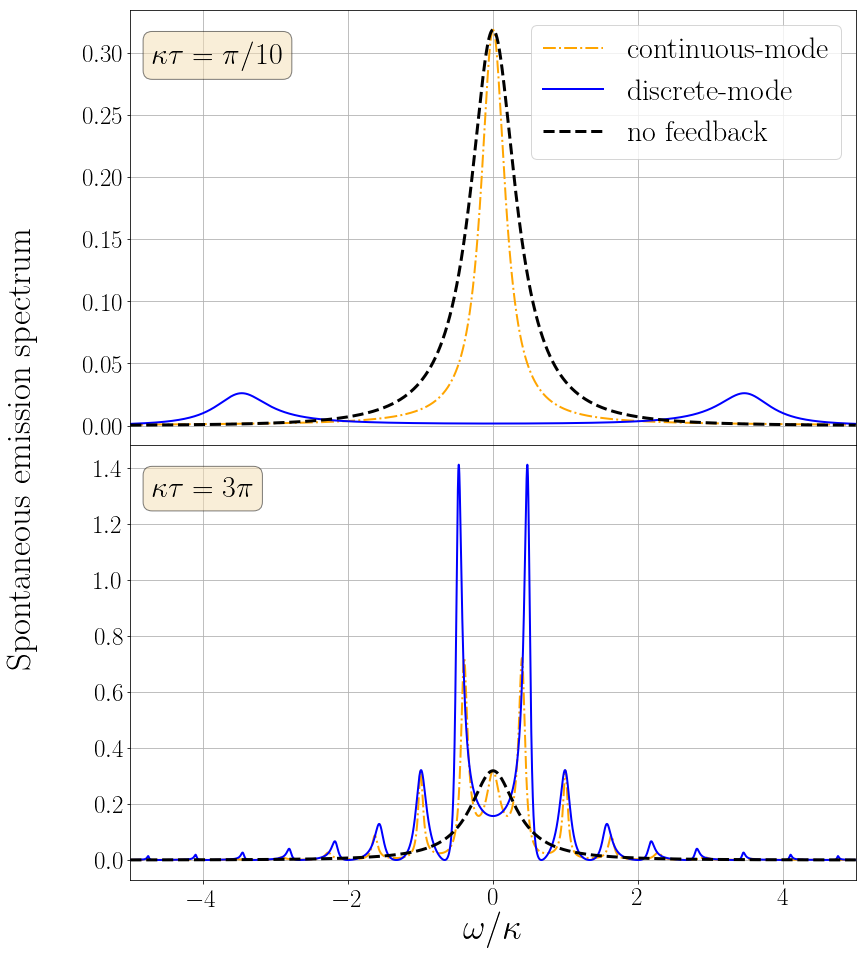}
\vglue -.5 cm
\caption{Spontaneous emission spectra detected through the loss channel described by $\kappa_1=\kappa/2$. The coupling strength is $\gamma=\kappa$ and the phase shift is $\phi=\pi$. \label{fig:spectrum}}
\end{figure}

\paragraph{Without feedback} In this case there is only a central peak broadened by the different loss channels:
\begin{align}
S^{(nofb)}(\omega) = \frac{2\gamma^2\kappa_1/\pi}{\lk\omega^2-\gamma^2\rk^2+4\lk\kappa_1+\kappa\rk^2\omega^2}.
\end{align}
\paragraph{Continuous-mode feedback} Considering the output channel characterized by $\kappa_1$ for the continuous-mode feedback case, we obtain the following analytical expression for the spontaneous emission spectrum:
\begin{align}
    S^{(CM)}(\omega) &= \frac{2\gamma^2\kappa_1}{\pi}\lka\lsz\omega^2-\gamma^2+2\omega\kappa\sin{(\omega\tau-\phi)}\rsz^2\right.\nn\\
    &\hspace{.3cm}+\left.4\lsz\kappa_1+\kappa-\kappa\cos{(\omega\tau-\phi)}\rsz^2\omega^2\rka^{-1}.
\end{align}

\reffig{fig:spectrum} shows how the spectra for various setups compare. In the upper panel the short delay case shows clear signatures of excitation trapping on resonance for a discrete mode structure. In the continuous-mode case there is a pronounced linewidth-narrowing compared to the case without feedback, which corresponds to a suppressed decay.

As the feedback delay is increasing, more and more frequencies contribute to the emission as shown in the lower panel in \reffig{fig:spectrum}. However, a reduced emission on resonance can still be observed for the discrete-mode case, which is what we expect from \refeq{eq:SS_trapped}. This emphasizes the main difference between the two cases, i.e. the discrete-mode case resembles a closed system, whereas the continuous mode structure results in an inherent dissipative nature of the feedback loop.

\section{Conclusion}
In this paper we demonstrated that whether continuous or discrete modes are considered in a coherent feedback setup is an important characteristic that essentially determines the qualitative behaviour of the overall scheme. The main difference lies in the open or closed system characteristic of the control setup.

The discrete set of environmental modes can be interpreted as a multimode cavity arrangement, which therefore enables coherent evolution towards excitation trapping in dark states. As this setup is primarily closed, no change can be introduced by the feedback to the original stability landscape of the system.

The other design using a continuous spectrum of reservoir modes shows open system characteristics, where no excitation trapping is possible. However, this single-delay coherent feedback can function well for stabilizing intrinsic quantum processes of the system such as Rabi oscillations in the Jaynes-Cummings model.

The results presented in this paper are of fundamental importance for the implementation of coherent feedback control schemes as they restrict the experimental designs to obtain a more targeted and thus improved performance.  As mentioned before in the introduction, both schemes are of relevance in terms of the considered applications. However, each targeted function has a preferred realization that should be taken into account. 

\acknowledgements
N.N. and S.P. are grateful to the group of A.K. in Berlin for the support and hospitality. We thank S. Hein for fruitful discussions in the early stage of the project. AK and AC gratefully acknowledge support from the Deutsche Forschungsgemeinschaft (DFG) through the project B1 of the SFB 910 and from the European Union’s Horizon	2020 research and innovation program under the	SONAR grant agreement no. [734690].

\appendix
\section{Contribution of \texorpdfstring{$\sin{(k_qL)}$}{Lg}\label{app:sinkL}}
Due to the definition of modes outside the cavity, we have the following expression:
\begin{align}
    \sin{(k_qL)} &= \sin{\lsz \frac{(2q+1)\pi}{2}+\theta_q\rsz}\nn\\
    &=\sin{\lsz\frac{(2q+1)\pi}{2}\rsz}\cos{\theta_q}\nn\\
    &= (-1)^q\cos{\theta_q} \approx(-1)^q,
\end{align}
the square of which is 1.

\begin{figure}[b!]
\centering
\includegraphics[width=0.48\textwidth,trim={0 0 0 0},clip]{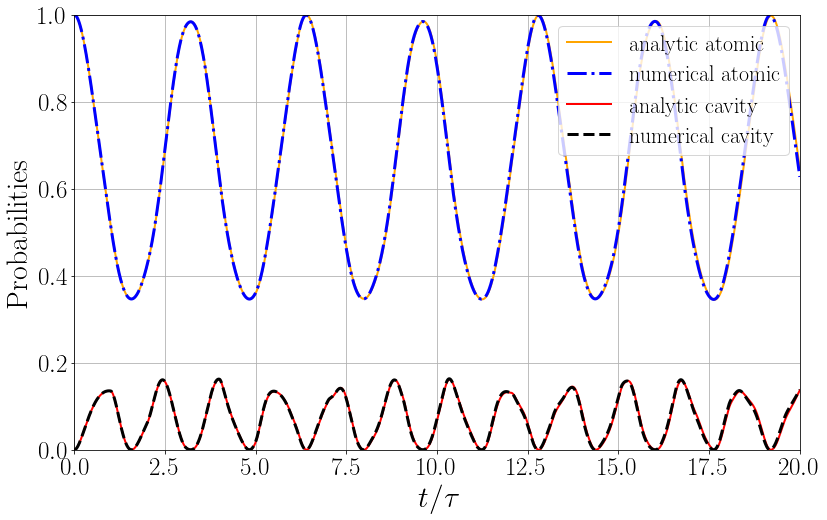}
\vglue -.5 cm
\caption{Comparison between the numerical, time-local solution for the cavity and atomic excitation probabilities and the evolution by delay differential equations. $\kappa\tau=\gamma\tau=\pi/3,\phi=\pi$. \label{fig:comparison}}
\end{figure}

\section{Comparison between Dirac comb and wave number sum}
Based on the main text there are two ways to simulate the dynamics of the system coupled to the discrete set of modes. One is following the time-local equations of motion in (\ref{eq:c_e_eq_MD}-\ref{eq:c_gqp_eq_MD}); in this case the number of considered modes has to be large enough so that it covers the whole relevant range of frequencies.

The other method involves solving the delayed differential equation \refeq{eq:cg_EoM} instead of \refeq{eq:c_g_eq_MD} and \refeq{eq:c_gqp_eq_MD}, thus directly incorporating the influence of the discrete environment. In \reffig{fig:comparison} we show the time evolution obtained by the two simulation methods. According to that, there is a good agreement between the two approaches.

\section{Long-time solution in the weak coupling regime}
When $\gamma=\kappap$, the time-evolution of the system can be expressed by using the same tricks as in \cite{Kabuss2015}:
\begin{align}
\label{eq:analytic}
&c_g^{(\infty)(DM)}(t) = \\
&\hspace{.2cm}=i\lka\sum_{m=1}^\infty\sum_{l=0}^m\sum_{p=0}^\infty(-4)^m(-1)^p
\begin{pmatrix}m\\l
\end{pmatrix}
\begin{pmatrix}p+m-1\\p
\end{pmatrix}
\nn\right.\\
&\left.\hspace{.9cm}\cdot\frac{\lsz\kappa(t-p\tau)\rsz^{m+l+1}}{(m+l+1)!}e^{\kappa(t-p\tau)-ip\phi}\Theta(t-p\tau)+\kappa te^{\kappa t}\rka\nn.
\end{align}

Let us compare the above result with the continuous-mode case, which was given in \cite{Kabuss2015}:
\begin{align}
&c_g^{(\infty)(CM)}(t) = \\
&\hspace{.2cm}i\lka\sum_{m=1}^\infty\sum_{l=0}^m2^m(-1)^l
\begin{pmatrix}m\\l
\end{pmatrix}
\frac{\lsz\kappa(t-l\tau)\rsz^{m+l+1}}{(m+l+1)!}\cdot\nn\right.\\
&\left.\hspace{.2cm}\cdot e^{-\kappa(t-l\tau)+il\phi}\Theta(t-l\tau)+
\kappa te^{-\kappa t}\rka.
\end{align}
The two expressions have many similarities with each other, however, notice the extra summation for the discrete-mode case which is due to the multiple recurring delay contributions. Also note that in this case each delayed term stabilizes the intrinsically unstable dynamics.
\bibstyle{apsrev4-1}
\bibliography{multidelay}

\end{document}